\documentstyle[12pt,aaspp4]{article}

\def\deg{\hbox{$^\circ$}}

\def\arcsec{\hbox{$^{\prime\prime}$}}
\def\fmag{\hbox{$.\!\!^m$}}

\begin{document}
\slugcomment{The paper published in {\it Bull. SAO. (Izv. SAO), 1993,
36, 5.}}


\title{SPECTRAL INVESTIGATIONS OF OBJECTS FROM THE SECOND
BYURAKAN SURVEY. STELLAR OBJECTS. VI\footnotemark
\footnotetext{Table 2 and Figures 1-25 see in original
paper Bull. SAO. 1993, 36, 5.}.}


\author{J.A. Stepanian and V.A. Lipovetsky}
\affil{Special Astrophysical Observatory RAS\ Nizhnij Arkhys,
Karachai-Cherkessia, 357147\\ Russia \\
Electronic mail: jstep@sao.ru}

\author{V.H. Chavushyan, L.K. Erastova, and S.K. Balayan}
\affil{Byurakan Astrophysical Observatory, Byurakan, 378433 Armenia \\
Electronic mail: vahram@bao.sci.am, lke@sao.ru end sbalayan@bao.sci.am}

\begin{abstract}

 The data for 100 QSOs from the Second Byurakan Survey (SBS) located on
the northern sky $8^{h} < \alpha < 17^{h}$ and $49\deg < \delta < 61\deg$
are presented. The main parameters of emission lines and the other data
for all QSOs as wel as their scans are given. Totally we have found 216
QSOs in SBS.

\end{abstract}

\section{INTRODUCTION}

 This is a sixth paper of this series. In the first five papers (Stepanian
et al. 1990a, b, c; 1991a, b) we have presented spectral data for 156
stellar objects, located in six fields of the SBS, and for 47 stellar
objects, located beyond the published SBS fields in the sky region
$8^{h} < \alpha < 17^{h}$ and $49\deg < \delta < 61\deg$. In the mentioned
region we have discovered 116 QSOs, one LINER, seven Sy1 galaxies, one Sy2
and five emission-line galaxies as wel.

 In this paper we present spectral and other data on the next 100
quasistellar objects. Three objects SBS~1010+465, SBS~1107+487 and SBS
1307+462, located in the zone bordering on the SBS region, are also
listed.

 All spectra have been obtained at the 6 m telescope with the 1024-chanel
TV scaner (IPCS) in the spectral range 3200-7100 \AA\AA, with a spectral
resolution of 10-15 \AA\ and the diafragm 1.5-2$\arcsec$. One or two
standard stars were observed every night exluding the bad weather
conditions. The method of observations, sampling, classification and data
processing are presented by Stepanian et al. (1990a).

\section{DISCUSSION THE RESULTS}

 In Table 1 we collected together the data on the investigated objects:
1 -- SBS designation, 2 and 3 -- equatorial coordinates for epoch 1950,
these are accurate to about $\pm 1\arcsec$, 4 -- visual estimates of
apparent magnitude in the blue band (B), 5 -- mean value of the redshift
determined due to the strong emission, 6 -- date of observations.

Table 2 presents the measuremant results of emission line parameters for
QSOs: 1 -- SBS designation, 2 -- visual estimate of the apparent magnitude
in blue band (B), 3 -- absolute magnitude calculated using equation from
Schmidt \& Green, 1993) for $H_{0}=50~km~s^{-1}~Mpc^{-1}$, $q_{0}=0$
and $\alpha=-0.7$:

\begin{equation}
\label{absmag1}
M(B)=B-5\log(z)(1+z/2)+2.5(1+\alpha)\log(1+z)-43.89
\end{equation}


4 -- mean value of the redshift determined due to the strong emission
lines, 5 -- observed wavelength of emission line, 6 and 7 -- laboratory
wavelength of emission line of ion and the ion, 8 -- total width of
emission line at the level of continuum $FWOI (km/s)$, 9 -- total width of
emission line at half intensity (maximum) $FWHM (km/s)$, 10 -- observed
value of equivalent width of emission line $EW_{obs}$(\AA).

 In a few cases depending on redshifts, some emission lines located near
the end of the spectra, where the spectra were too deep and we could not
fit the right level of continuum, therefore for these objects the values
of $FWOM$, $FWHM$ and $EW$ in fact are more uncertain than for others,
for some of them we can't give their parameters.

 We have given the mean values for the quantities listed in Table 2, when
several spectra for one object are available. The sign colon implies
uncertain determination of a parameter. It can be see from Table 2 that
the range of apparent magnitudes $15\fmag5 \le m_{B} \le 19\fmag5$. The
range of redshifts of QSOs is $0.150 \le z \le 3.150$, luminosities are in
interval $-23\fmag3 \le M_{B} \le -31\fmag1$. In Figs. 1-25 the scans of
all QSOs are presented. The finding charts for all objects will be
published later in general SBS Catalogue.

\section{SHORT REMARKS ON SOME OBJECTS OF TABLE 1 AND 2}

\begin{description}

\item[0806+505] -- Very narrow $CIII]$ ($\lambda_{0}$~1909\AA).

\item[0817+573] -- At $\lambda_{obs}$~4190\AA\ and $\lambda_{obs}$~4715\AA\
the lines $HeII$ ($\lambda_{0}$~1640\AA) and ($\lambda_{0}$~1858\AA) are
suspected.

\item[0817+573] -- At $\lambda_{obs}$~3590\AA\ absorption detail is present.

\item[0929+521] -- $Ly_{\alpha}$ emission line consist of two narrow and
wide components.

\item[0949+527] -- Absorption detail at $\lambda_{obs}$~4285\AA\ is
present.

\item[1052+518] -- At $\lambda_{obs}$~6675\AA\ $MgII$
($\lambda_{0}$~2798\AA) is suspected.

\item[1107+487] -- The object is given also by Sanduleak \& Pesch (1989b)
where the value of $z_{em}\approx3.0$ was astimated from low-dispersion
objective prism spectrum.

\item[1116+610] -- In Markarian et al. (1984) the redshift value is wrong.

\item[1240+546] -- $CIII]$ ($\lambda_{0}$~1909\AA) emission line suspected
at $\lambda_{obs}$~6300\AA.

\item[1247+527] -- At ($\lambda_{obs}$~6820\AA) $HeI$
($\lambda_{0}$~5876\AA) emission line is suspected.

\item[1303+532] -- Absorption detail is suspected
at $\lambda_{obs}$~4370\AA).

\item[1307+462] -- $Ly_{\alpha}$ emission line consist of two narrow
and wide components.

\item[1308+512] -- The profiles of $Ly_{\beta}+OVI$ and $Ly_{\alpha}+NV$
are strongly blended by absorption details. Numerous absorption lines are
observed on the shortwave of $Ly_{\alpha}$. A strong absorption detail
with $FWHM\approx4000~km/s$ is observed at $\lambda_{obs}$~3800\AA.
Probably damped $Ly_{\alpha}$ absorption system. The spectrum is
underexposed, therefore the line parameters are defined uncertainly.

\item[1342+560] -- The spectra are not corrected for the spectral
sensitivity.

\item[1349+575] -- The profiles of lines $Ly_{\beta}+OVI$ and
$Ly_{\alpha}+NV$ are strongly blended by absorption details. Numerous
absorption lines are observed on the shortwave of $Ly_{\alpha}$.
An atention is drawn by two strong absorption details at
$\lambda_{obs}$~3745\AA\ with $FWHM\approx5000~km/s$ and
$\lambda_{obs}$~4460\AA\ with $FWHM\approx6000~km/s$. Probably BAL QSO,
atherwise damped $Ly_{\alpha}$ absorption system QSO.

\item[1406+492] -- The profiles of $CIV$ are strongly blended by absorption
details. CSO 609 (Sanduleak \& Pesch, 1989a).

\item[1408+544] -- $CIV$ line has strong absorption in the middle. Strong
absorption details are observed in the high resolution spectrum from the
shortwave of $Ly_{\alpha}$.

\item[1417+596] -- Absorption details, shifted from the emission lines at
velocities of 6800, 600 and 5800 $km/s$, respectively, are observed in the
blue wing of $Ly_{\alpha}+NV$, $SiIV+OIV]$ and $CIV$ emission lines. $FWHM$ of
absorption line the blue wing of $Ly_{\alpha}\approx3300~km/s$. Probably
BAL QSO. The line profiles are strongly blended by absorption details.
therefore the line parameters have been determined uncertainly.

\item[1419+538] -- $Ly_{\alpha}+NV$ line is on the edge of the spectrum,
therefore the lower values of the line parameters are presented.

\item[1421+511] -- CSO 643 (Sanduleak \& Pesch, 1989a).

\item[1424+502] -- Possible identification $z_{em}=1.553$.

\item[1426+506] -- CSO 654 (Sanduleak \& Pesch, 1989a).

\item[1437+509] -- CSO 677 (Sanduleak \& Pesch, 1989a).

\item[1439+522] -- CSO 680 (Sanduleak \& Pesch, 1989a).

\item[1500+557] -- Possible identification $z_{em}=0.475$.

\item[1504+543] -- CSO 722 (Sanduleak \& Pesch, 1989a).

\item[1518+497] -- The identification $z_{em}=1.6$ is also possible.

\item[1526+540] -- CSO 758 (Sanduleak \& Pesch, 1989a).

\item[1527+522] -- CSO 759 (Sanduleak \& Pesch, 1989a).

\item[1532+588] -- In the paper Stepanian et al. (1986) this object has
the number 55, where the wrong coordinates are presented.

\end{description}

\section{CONCLUSIONS}

 Among 100 investigated quasistellar objects we discovered one candidate
for damped $Ly_{\alpha}$ absorption system -- SBS~1308+512, and two BAL
QSOs -- SBS~1349+575 and SBS~1417+596. It should be noted that in two
remanining objects the absorption features have intermadiate $FWHM$ between
those of BAL QSOs and  $Ly_{\alpha}$ forest.

 So, the data of spectral studies of 216 QSOs from SBS survey are
presented.

\section{ACKNOWLEGEMENTS}

 The autors thank their collegues A. Knyazev, A. Ugryumov and S.
Pustil'nik for help during observations. All observations were fulfiled
due to observational time provided by the 6 m Telescope Time Allocation
Comission.


\begin{center}
{\bf Figure Captions}
\end{center}

\figcaption{ Figs 1-25. Plot the spectra of the observed QSOs in relative
flux units. Wavelengths in angstroms.}

\end{document}